\input{aipcheck}

\documentclass[,final] {aipproc}
\layoutstyle{8x11single}
\begin{document}
\title{Quasars and Galactic Nuclei,\\ a Half-Century Agitated Story}
\author{Suzy Collin}{  address={LUTH, Observatoire de Paris-Meudon, 92195, Meudon, France}}

\maketitle
\section{Motivation} My intention in this talk is to share a story in which I was very fortunate to participate almost since the beginning, but which is often ignored by the young generation of astronomers. It is the story of the discovery of Massive Black Holes (MBHs). Since everybody in this assembly knows well the subject in its present stage of development, I thought indeed that it could be interesting to show how the ideas that people take for granted presently had such difficulties to emerge and to gain credence. I think that this subject allows, better than any others, to observe that research is not ``a long quiet river'', and on the contrary evolves in a non-linear and erratic way,  full of mistakes and of dead ends, and that it gives rise to passionate controversies. We will see that the story of MBHs is made of fruitless researches opening on unexpected discoveries, come-backs of visionary models which were first neglected, temporary very fashionable but wrong models, strong debates involving even new physical laws, misinterpretations responsible for decades of stagnation, thousands of papers and nights of the largest telescopes.
	But finally it opened on a coherent physical model and on a new vision of galaxy evolution. 
	Since it is a long story, I have selected only a few fragments.
	
 \section{Prehistory} The story began in the forties, with a paper by Seyfert \cite{Seyfert1943} called simply ``Nuclear emission in spiral nebulae'' (at this time it was still usual to call a galaxy a ``nebula''), where he studied 6 galaxies characterized by a bright stellar-like nucleus (as bright as the whole galaxy in some cases), blue, and presenting all broad emission lines, some being of high excitation. He attributed the widths of the lines to Doppler motions, which could be as high as 8500km/s. This paper was never referred to during 16 years after its publication. 
 
	The story went on with the development of radio-astronomy, the theory of the synchrotron process, and the discovery that the strongest radio-sources, like the Crab, were emitting synchrotron  radiation. In 1954, Baade and Minkowski \cite{Baade1954} succeeded in identifying one of the most intense radio-source with a remote galaxy, Cygnus A. They identified also M87, the central galaxy of the Virgo cluster which displays a beautiful blue jet known since the beginning of the century, and several other galaxies with extended radio-sources. Burbidge \cite{Burbidge1959} thus concluded, on the basis of simple considerations of energy equipartition, that these sources must contain an enormous energy in magnetic field and relativistic particles, at least of the order of that obtained by the complete transformation of 10$^6$ to 10$^8$ solar masses into energy (the uncertainty comes the fact that one does not know how many protons are associated with the relativistic electrons). This conclusion  did not raise a strong interest in the community as it should have done.
	
	In 1959, Seyfert galaxies were exhumed because some visionary people realized that what occurs in their nucleus could have a relation with the phenomenon taking place in radio-galaxies. It was the beginning of something which nobody anticipated that it would become perhaps the strongest controversy in astronomy, lasting for twenty years (and even lasting now).  Burbidge, Burbidge \& Prendergast \cite{Burbidge-Prendergast1959} on one side and Woltjer \cite{Woltjer1959}, on the other side, published in the same issue of Astrophysical Journal but completely independently a paper about Seyfert galaxies. They came to exactly opposite conclusions.  Burbidge et al. studied the rotation curve of the Seyfert galaxy NGC 1068, and deduced that the plasma giving rise to the emission lines should be ejected from the nucleus, while Woltjer, discussing the properties of the line emission in all six Seyfert galaxies, concluded that it should be gravitationally confined by a massive body. Woltjer was right, but with wrong arguments, because he overestimated the mass of the ejected gas and concluded that it could not be ejected unless the whole galaxy would have been emptied too rapidly. Burbidge et al. were right because NGC 1068 is a peculiar Seyfert (we call it now a Seyfert 2), whose emission lines are formed at a much larger distance than in the other Seyfert galaxies. 
	
	Few weeks before the publication of the discovery of quasars, Burbidge, Burbidge \& Sandage \cite{Burbidge-Sandage1963} published a big paper called ``Evidence for the Occurrence of Violent Events in the Nuclei of Galaxies'' which was a prefiguration of quasars from the energetic point of view and should have opened the way for these objects - but it did not.

\section{Quasar discovery}

	Optically identifying the radiosources of the third Cambridge catalogue led to the discovery of a class of sources without any visible counterpart such as galaxies or supernovae remnants. At the position of the source number 48 laid a blue star in which a spectrum obtained with the Palomar telescope revealed several intense and broad emission lines at unknown positions. This was announced at the American Astronomical Society meeting in December 1960, but probably because the astronomers did not know what to do with this strange spectrum and thought that the identification of the source was wrong, they did not publish this result. In 1962, there were two occultations by the moon of the source number 273 visible in Australia, allowing Hazard to obtain a very precise position of the object. At this position laid also a blue star. A spectrum was immediately taken by Schmidt at Palomar, and it revealed intense broad emission lines, again unknown. The breakthrough came on February 5, when Schmidt realized that the lines were the Balmer lines H$\delta$, H$\gamma$ and H$\beta$, but redshifted by 16$\%$. Four papers were published in the same issue of Nature about the discovery of 3C273 (\cite{Hazard1963}, \cite{Schmidt1963}, \cite{Oke1963}, \cite{Greenstein1963}). Immediately Greenstein and Mathews pulled out of their drawers their old spectrum of 3C48, and were  able to identify an ultraviolet line of MgII and the Balmer lines, all redshifted by 37$\%$. Soon after, two other sources were also identified with a high redshift, and it was decided to held a meeting in Dallas to discuss this astonishing discovery. It was followed the year after by another in Austin. These were the two first Texas symposia, which now take place every two years (but no more in Texas!). Immediately after, several summer schools were devoted to the subject, and several research or popular books were published about these  ``quasi stellar objects'', which soon became simply ``quasars''. Considering the importance of the discovery, the subject was not much discussed during the general assembly of International Astronomical Union in Hamburg in August 1964. Apparently something was disturbing ``classical'' astronomers. An interesting question to ask also is whether quasars would have been identified so rapidly if by chance one of the most powerful of them, 3C273, would not have had a small redshift, so that its line spectrum looked familiar to people used to observe planetary nebulae. 
			
	Completely unforeseen was the discovery by Sandage \cite{Sandage1965} of a new class of quasars, this time radio quiet. It soon became clear that they were much more numerous than radio loud quasars. It appeared also that there were a whole bunch of objects resembling the quasars, which were called according to their appearance Opically Violently Variable, Highly Polarized Objects, BLac objects, etc. In spite of many differences, all these objects shared common properties: they were all very small, with a variable continuum extending over six up to twelve decades of frequency (but the latter only for some radio-sources). It led to the idea that a similar mechanism was at work in all these objects. The word ``Active Galactic Nuclei'' (AGN) was invented  for Seyfert galaxies, and rapidly one was led to the idea that quasars were sitting inside nuclei of giant remote galaxies, and that they were some kind of gigantic version of AGN. But this was proved to be the truth only twenty years later, when the first images of the ``host-galaxies'' of quasars were obtained. 
	
\section{The redshift controversy}

The first problems appeared immediately. 
In 1964 Greenstein and Schmidt \cite{Greenstein1964} made a thorough analysis of the results concerning 2C273 and 3C48 and concluded that the only possible explanation for the redshifts was the Hubble law. Gravitational redshift was indeed definively eliminated on the basis of the width of the lines. This result put the objects at distances of one to several billions light-years, which meant that they were very luminous, typically 10$^{46}$ ergs s$^{-1}$ in the visible band, i.e. as much as 1000 galaxies like ours. But on the other hand it was realized that 3C273 was a star-like object observed since a century by the Harvard Observatory and classified as being variable, with sometimes variations by one magnitude in a week. So this incredibly powerful radiation was emitted in a region smaller than a light-week! Then, it was discovered that one of its two radio components was varying on a time scale of a year. When Sciama received this news during the Varenna school on High Energy Astrophysics (1965)  - which I was attending -  he changed the subject of his lectures to discuss the problem, and found the explanation (one component was self-absorbed and had a very small dimension).  All this was raising very difficult problems, and some people argued that the cosmological interpretation of the redshift was wrong. But what other interpretation to give? It could not be Doppler redshifts, since there should have been as many blueshifts as redshifts, while at this time one knew already several of these objects, and they all had redshifts. So it was necessary to invoke a still unknown physical law to explain the redshifts. 

This started the redshift controversy, which  lasted during about 15 years, and  occupied a large fraction of  the meetings during all this time. The discoveries concerning quasars were indeed so unprecedented, and not immediately understandable, that they permanently provoked hard debates, and created the idea that something else than the known physical laws was at work. 

\begin{figure} 
\includegraphics[height=.3\textheight]{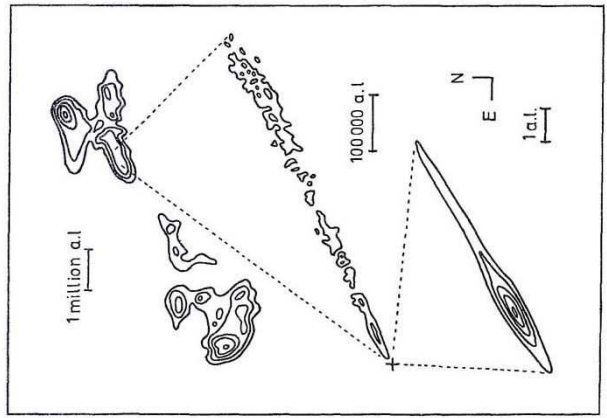}  
\caption{This radio map of NGC 6251, as published by Readhead, Cohen \& Blandford in 1978, shows that a small jet 5 light-years long is aligned with a larger jet of 600 000 light-years, itself aligned with the direction of the radio lobes, separated by 9 millions light-years. The fact that the two jets at the small and intermediate scales are seen only on one side, while the lobes at large scale are almost symmetrical with respect to the galaxy, proves that the side of the jet directed towards us is relativistic boosted, and therefore that the bulk velocity of the jet is very close to the velocity of light. }  
\label{poupees-gigognes}
\end{figure}

One paradox was linked to the very small size of the radio-source. This small size was deduced from the fact that the synchrotron radio spectrum was self-absorbed, like in 3C273. Hoyle, Burbidge \& Sargent \cite{Hoyle1966} noticed that this implies a very high density of radiation, meaning that the relativistic electrons themselves should hit the radio photons and transform them by inverse Compton process into visible, then X, then gamma ray photons (this is the well-known Synchrotron-Self-Compton mechanism, which accounts for the X-ray and gamma ray emission of radio loud quasars and compact sources in radio galaxies). This process would cause the rapid disappearence of the radio source, and it would produce intense gamma ray sources\footnote{It is worth noting that such a mechanism is at work in blazars, which are  relativistically boosted, and where the gamma-ray band contains sometimes more energy than the radio and the X-ray bands.}. It was called the ``Compton catastrophy''. But it was possible to overcome the problem by assuming simply that the source was located in our Galaxy, because in this case the density of radiation and relativistic electrons would have been much lower. This thesis was defended by the proponents of non-cosmological redshifts. Woltjer \cite{Woltjer1966} first replied that Compton electrons are emitted preferentially in the direction of the electron velocity, which amplifies the radiation, and that the number of collisions between electrons and photons is reduced. Then Rees, who was at this time a PhD student of Sciama, found the solution of the paradox by invoking relativistic motions \cite{Rees1967}. In the same paper he also predicted that ``an object moving relativistically in suitable directions may appear to a distant observer to have a transverse velocity much greater than the velocity of light''. This was indeed observed a few years later, but Rees' explanation was forgotten at this time! With the advent of Very Large Baseline Interferometry (VLBI), a radio-source moving away apparently at a speed larger than the light velocity was discovered in 3C279, then in 3C273 and other extragalactic radio-sources. But the transverse velocity was large only because the distance of the object was assumed to be cosmological: if the object would be local, this would not be the case. So there were lots of discussion on these phenomena in the framework of the redshift controversy. Several models were invoked to solve the paradox of these ``superluminic sources'' , as they were called, and reconcile them with the cosmological hypothesis; in particular the possibility of a phase velocity, or erratic light flashes like on a Christmas tree. But further observations showed that the phenomenon was not erratic, and that the ``blobs'' of light were going always in the same direction, as if they were ejected from a central object. Everybody knows presently that the problem is easily solved, owing to the finite light velocity, by assuming simply a source ejected from the center at a speed close to the light velocity and seen at a relatively small inclination (actually not very small, it depends on the Lorentz factor). But there was no proof at this time that the bulk motions were relativistic like the erratic motions of the synchrotron electrons. 	A fundamental discovery occurred then, which proved the correctness of the argument, and at the same time definitively showed  that the inner parsec of the nucleus was at the origin of the relativistic jet and of the enormous amount of energy stored in the giant radio lobes sitting around radio-galaxies: it was a radio map of the galaxy NGC 6251 published in 1978  by Readhead et al. \cite{Readhead1978} in the proceedings of the Copenhagen meeting (cf. Fig. \ref{poupees-gigognes}).

A prediction of the cosmological hypothesis was that one should observe a continuous absorption on the blue side of the L$\alpha$ lines of quasars, due to intergalactic neutral hydrogen  less redshifted than the quasar emission line on the line of sight. It was called the ``Gunn-Peterson test'' \cite{Gunn1965} (though M. Burbidge also proposed the same year to try to detect this absorption). This absorption was never detected though it was intensively looked for, and this was used as an argument against cosmological redshifts (it could also be due to the fact that intergalactic hydrogen was absent, or completely ionized). But a completely unexpected observation occurred instead. Burbidge, Burbidge \& Lynds \cite{Burbidge-Lynds1966} discovered several narrow absorption lines in the spectrum of the quasar 3C191, lying at the blue side of the emission lines. They were immediately attributed to an outflow coming from the quasar and moving towards us. A year after, this explanation was reinforced by the discovery by Burbidge \& Lynds of absorption lines in the blue wing of another quasar, PKS 5200 \cite{Lynds1967}, but this time broad with the classical P Cygni profile, which is the signature of a wind outflowing from a central body. In the following years, many systems of absorption lines were discovered in other quasars. A big problem appeared then: some of these systems observed  in high redshift quasars had low redshifts, meaning that the relative velocity of the wind was close to the velocity of light. This was already embarrassing; but an even worse problem was that the absorption lines were narrow, implying an extremely small velocity dispersion (a few tens km/sec) in a plasma accelerated up to almost the light velocity. Several theoricists, and not the least ones, did not succeed in explaining such a phenomenon. There was however a very simple explanation: these narrow absorption lines were due to clouds located on the line of sight of the quasars, most probably in intervening galaxies. But this explanation was refused by the ``non-cosmologist'' communauty until the end of the seventies, partly because galaxies were not supposed to be as numerous as they are in reality at high redshifts, and because they were also not assumed to extend as far as they do. This explanation was finally accepted when Bergeron \cite{Bergeron1986} and others were able to detect galaxies near the line of sight of the quasars, having exactly the same redshift as the absorption systems. 

 So the redshift controversy ended abruptly at the beginning of the eighties, at least for a large majority of quasar astronomers. During all these years, several examples of apparently strange regularities in the redshift distribution, groups of galaxies containing high redshift quasars, alignments of quasars, etc., have been permanently produced by Arp. Some of them - but not all, because it would take centuries! - were discussed and were showed to be due to chance or to observational bias. Strong battles took place in several meetings. And still now, some very well-known people, such as M. and G. Burbidge, Narlikar, Arp, are supporting the  non-cosmological hypothesis. Their ideas have evolved with time, but they still believe that a fraction of redshifts are ``anomalous'', i.e. due to an unknown physical law, and that quasars are ejected by local AGN (for instance in couples of discrepant redshifts, like NGC4319 and Mrk205, or in the nearby galaxy NGC 7320 which contains a z=2 quasar, according to Galliani et al. \cite{Galianni2005}).

\section{The central engine}
\begin{figure} 
\includegraphics[height=.3\textheight]{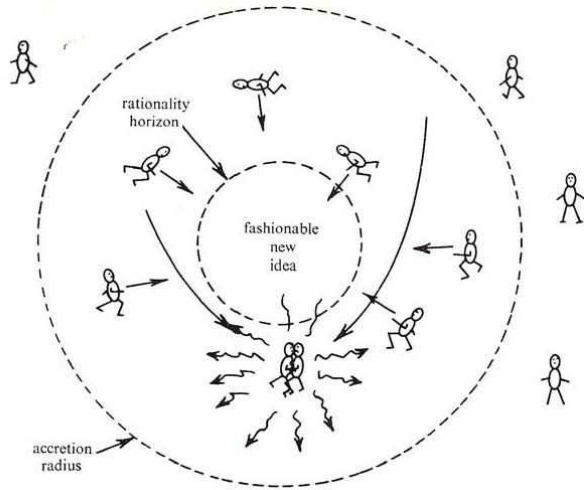}  
\caption{Cartoon produced by McCray at the Cambridge summer school in 1977 and called ``Response of astrophysicists to a fashionable new idea''. I extract a few lines from his paper: ``Beyond the accretion radius, $r_a$, astrophysicists are sufficiently busy to not be influenced by the fashionable new idea. But others, within $r_a$, begin a headlong plunge towards it.... In their rush to be the first, they almost invariably miss the central point, and fly off on some tangent.... In the vicinity of the idea, communication must finally occur, but it does so in violent collisions.... Some individuals may have crossed the rationality horizon $r_s$ beyond which the fashionable idea has become an article of faith. These unfortunate souls never escape. Examples of this latter phenomenon are also familiar to all of us.''}  
\label{mcCray}
\end{figure}
Since the beginning, the most important problem was to understand the origin of the energy in quasars. It took about fifteen years, though some people have found immediately the correct explanation. Salpeter  \cite{Salpeter1964} and Zel'dovich \cite{Zeldovich1965} suggested independently that a Massive Black Hole (MBH) was present in these objects, and Salpeter proposed that the matter and the angular momentum transport necessary for accretion was accomplished via a turbulent viscosity (this is the presently most accepted view). But nobody took this idea seriously. Black Holes as they were named by Wheeler in 1967 were rapidly quite popular among theoritical physicists, and a large litterature was devoted to them, but they were considered as an utopia by astronomers, and not associated with the release of energy in AGN. Then Lynden-Bell  published in 1969 a premonitory idea \cite{Lynden-Bell1969}: ``Galactic Nuclei as Collapsed Old Quasars'', and reiterated the proposition at the famous Vatican Conference in 1970 on ``Nuclei of Galaxies''\footnote{These meetings in the Vatican take place from time to time and I think it was the first Vatican meeting devoted to astrophysics. They gather a few of the best scientists in a given science, to discuss some fundamental issue on which experts have contradictory opinions.}. Though 25 astronomers among the most well-known were attending it, apparently nobody realized that this model involving a black hole could be interesting. Actually many mechanisms were discussed in the sixties, like chain reactions of supernovae, coeval evolving star cluster, ``galactic flares'', stellar collisions, but the most popular was unquestionably supermassive stars: I think the reason was because stellar studies had been very successful during the previous decades, and some of their best promotors (the Burbidges, Hoyle, Fowler), were active in the quasar and AGN field. But all these mechanisms were dismissed for one reason or another: instabilities for supermassive stars, inefficiency for others, or inability of accounting for rapid changes of luminosity. A funny thing is that the MBH idea was helped by the discovery of very rapid and strong flux variations in BLac and Blazars, while it is well known now that they are relativistically boosted and that the very strong variations of luminosity are only apparent!

Finally M. Rees produced in 1977 at the Copenhagen meeting what he called ``the flow chart'' of a galactic nucleus, where he showed that its fate is to lead inevitably to the building of a MBH through several different ways, in a time scale smaller than the Hubble time \cite{Rees1978}. Since MBHs were now considered in a completely realistic astrophysical environment, more and more people signified their adhesion to the model.
Though he was himself believing in the MBH model, McCray produced a month later a well-known humorous cartoon at a summer school in Cambridge (\cite{McCray}), which was designed to illustrate spherical accretion, but in reality was used to illustrate the astrophysicists sociology. I cannot resist to show it (cf. Fig. \ref{mcCray}). 

The idea of MBH won also partly because of the discovery of pulsars in 1968, which showed that a very compact object like a neutron star could be a reality, and partly because the masses of the compact stars in X-ray binaries like Cygnus X1 were determined to be larger than the acceptable mass of neutron stars, so they should be necessarily stellar black holes. In a sense quasars were discovered too early, and the minds were not prepared to them, contrary to what happened for instance to the cosmological blackbody radiation which was already predicted and looked for when it was discovered (by chance!). 

Personally I think also that a very important discovery was that made by Shields \cite{Shields1978} when he identified a ``Big Blue Bump'' in the optical-UV spectrum of quasars, and showed that it is due to thermal emission of an accretion disk. Before that, during fifteen years, the whole continuum of quasars from the infrared to the X-ray band has always been  attributed to non-thermal processes (synchrotron, and synchrotron-self-Compton), owing to a rough similarity with a power law, and to strong prejudices coming from the radiosources. It is widely admitted now that the continuum emission of radio quiet quasars is entirely thermal and produced by the accretion flow.

	So we have seen that before the recognition of MBHs as the engines powering quasars and AGN, there were almost two decades of hesitations. After the acceptation of the MBHs, they became a paradigm, as well as the existence of the accretion disk. Thus many progresses were made simply because people knew what to look for, contrary to the previous years. 	
\begin{figure} 
\includegraphics[height=.3\textheight]{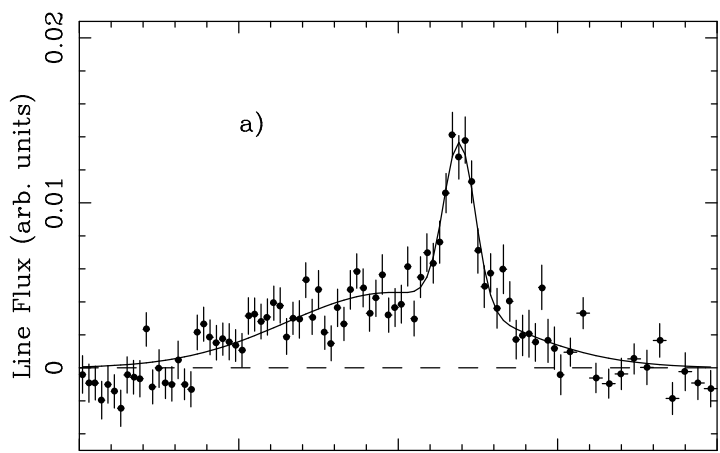}  
\caption{Composite line profile of the FeK line at 6.4 keV obtained by combining the data of 14 Seyfert 1 nuclei, showing the extended red wing due to gravitational redshift, from Nandra et al. \cite{Nandra1997}.  }  
\label{Fe-profile-Nandra97}
\end{figure}
A definitive proof of the presence of MBHs at the centers of AGN was the observation of an X-ray fluorescent iron K line at 6.4 keV. The profile of the line displayed indeed an extended red wing characteristic of gravitational redshift (cf. Fig. \ref{Fe-profile-Nandra97}). The line is interpreted as produced by the ``reflection'' on the accretion disk of an X-ray source located at a distance of the order of a few tens gravitational radii from the MBH. In one case it seems even that the line is formed at a distance of 2 gravitational radii, which means that the MBH is a rotating Kerr and not a Schwarzschild black hole (\cite{Young2005}). 

When black holes were discovered, it was obvious that objects like quasars and Seyfert nuclei should differ by the black hole masses and the accretion rates. But this was not sufficient to explain the great diversity of observed objects. Then Antonucci and Miller showed in 1985 \cite{Antonucci1985} that Seyfert 2 galaxies (different from Seyfert 1 by the absence of broad lines) were actually Seyfert 1 viewed ``edge-on'' through a kind of dusty outer disk, which was called ``the torus''. It amounted to recognizing the influence of the angle of view, which was quite expected, as it is due to the existence of the rotation axis of the accretion disk. (at least this time an expectation was correct!). The phenomenon was called the ``Unified Scheme'' of Seyfert nuclei, and it was extended to other objects. Barthel \cite{Barthel1989} suggested that radio-galaxies associated with strong radio-sources showing large radio-lobes (actually those which are called FRII) contained radio-quasars viewed through the dusty torus, and this idea was proved to be correct later. One consequence of the existence of  ``hidden'' quasars and AGN is that they contribute for a large fraction to the cosmological diffuse X-ray background. This solved a long-standing paradox concerning the shape of the diffuse X-ray spectrum.  Finally, since the end of the seventies, it was already admitted that BLac objects and Blazars (radio-loud quasars with a polarized and strongly variable continuum, and very compact radio cores) were seen almost in the direction of a relativistic jet and therefore relativistically boosted. Now every one knows that the direction of the jet is that of the disk axis. 

Another parameter, besides the viewing angle, the mass of the black hole and its accretion rate, can also be added to unification schemes of MBHs, namely the spin of the black hole. The spin is indeed sometimes invoked to explain the - still not understood - division between radio-loud and radio-quiet objects: radio-loud objects contain highly collimated relativistic jets which are not found in radio-quiet ones (where they seem to be replaced by winds, for a reason which is not clear), and these jets could be powered by a magneto-centrifugal mechanism anchored in a rotating black hole (Blandford-Znajek \cite{Blandford1977}). 

During the nineties, the idea that quiescent galaxies (i.e. with a non-active nucleus) could also harbor a MBH began to win popularity, though in the eighties a majority of people were reluctant to accept it.  The  discovery that ultraluminous infrared galaxies harbor dust enshrouded AGN, and are systematically in strong interaction with another galaxy, led to the idea that MBHs could be powered at least indirectly by galaxy interactions through gravitational torques. This was proved by numerical simulations. It showed that the environment of black holes had also to be taken into account. 

\section{Quasars in the cosmological context}	
When quasars were discovered, since they were the most remote objects in the Universe, they raised the great hope that some cosmological problem could be solved with their help. The first hope was to build a ``Hubble diagram'' of quasars in order to deduce H$_0$ and q$_0$. But for this quasars ought to  be ``standard candles'' distributed uniformly accross the Universe, which is far from being the case! Not only they span a large range of luminosities, but also they undergo a great cosmological evolution in luminosity and/or in number, as it was shown by Schmidt  already in 1967. So they had to be dismissed for cosmological tests. The second hope, also not fulfilled, was to measure the quantity of neutral hydrogen in the intergalactic space, as explained before. But curiously, the detection of the absorption systems, and in particular of the unexpected ``L$\alpha$ forest''  - a series of discrete L$\alpha$ absorptions due to clouds made of pure hydrogen, actually replacing the continuous absorption predicted by Gunn and Peterson - opened a new era for cosmological studies. 

		However the most important discovery in this context was  the (again unexpected) relation between the masses of MBHs and the masses of galactic bulges, implying a parallel cosmological evolution of MBHs and galaxies. This issue has been completely neglected until the end of the twentieth century. 
\begin{figure} 
\includegraphics[height=.3\textheight]{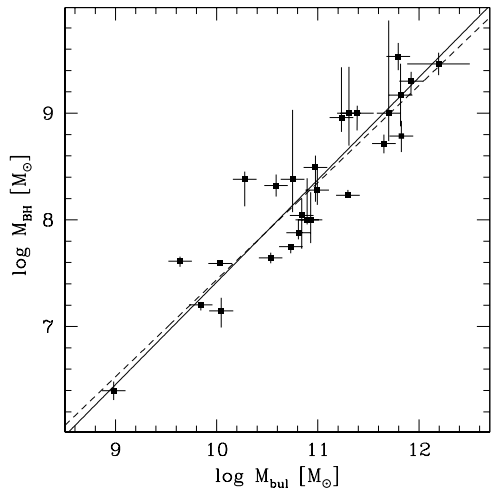}  
\caption{Relation between the masses of the central black holes and the mass of the bulges of the host galaxies (from Marconi and Hunt \cite{Marconi2004}).  } 
\label{MBH-vs-Mbul}
\end{figure}		
In 1982 Soltan \cite{Soltan1982} computed from the observation  of the luminosity function of quasars the mass density locked in MBHs, assuming simply that quasars were getting their power through accretion onto the black holes, and that they were radiating at the Eddington luminosity (the maximum allowed luminosity for an accreting object). His computation gave thus a minimum value for the co-moving density, and it was much larger than that of local MBHs in AGN. It meant that these MBHs had to lay somewhere else. Rees \cite{Rees1988} argued that they should be in the nuclei of quiescent galaxies. Through dynamical spectroscopical studies performed with the Hubble space telescope and with the largest ground based telescopes, the masses of the inner parsec in the nuclei of nearby quiescent galaxies were determined in the nineties. They were shown to harbor from 10${^7}$ up to 10${^9}$ solar masses, having a high mass / luminosity ratio; but it was not clear that these masses were in the form of MBHs (though the collapse time of a dense star cluster towards a MBH would have been smaller than the Hubble time). So this idea was not taken very seriously before the discovery of our own central MBH, obtained through the study of the motions of stars at very short distance from the putative massive black hole, SgA* (Genzel et al 1996 \cite{Genzel1996}). A non-ubiquitous MBH of 4 10$^6$ solar masses lies at the center of our Galaxy. 		

I think that this discovery impulsed the hypothesis of MBHs sitting in the nucleus of quiescent galaxies. The next step was the (again unexpected!) discovery of a relationship between the mass of these black holes and the luminosity of the bulges (Magorrian et al. 1998 \cite{Magorrian1998}),  followed by that of a tighter relationship between the masses of the MBHs and the dispersion velocity of the stars in the bulges, which is directly related to the  bulge masses (\cite{Ferrarese2000}, \cite{Gebhardt2000} ). Roughly, the mass of the MBH is equal to 2/1000 of the mass of the bulge (cf. Fig. \ref{MBH-vs-Mbul}), indicating a strong link between them. Now, the formation and the evolution of MBHs is systematically placed in the context of galaxy evolution, generally through cold dark matter hierarchical scenario.  Several explanations have also been proposed to account for the black hole / bulge mass relationship, in particular a feedback from the growing quasar. 		

What are the positions of AGN in the black hole / bulge mass relationship? The masses of black holes in AGN are determined in a completely different way as those of quiescent galaxies, using a technique called ``reverberation mapping''. It consists in measuring the velocity of the lines - which are assumed to be gravitationally bound to the black hole - and the time delay between the variations of the line and the continuum flux, from which is deduced the size of the emitting region. Actually, in my thesis in 1968 I argued that since the photoionizing continuum was variable, one should also expect in response variations of the broad lines in a timescale of a year. And indeed Andrillat and myself \cite{Andrillat1968} noticed that in the Seyfert galaxy NGC 3516, H$\beta$ was much less intense than in the original Seyfert spectrum; we confirmed the variability of this nucleus a few years later after having ourselves observed regularly this object. But at that time neither the prediction nor the observations were believed. Later on, intense monitoring campaigns to perform reverberation mapping of the BLR were organized, whose result was to lead recently to the masses of MBHs in several dozen of quasars and AGN. In the course of these studies, the existence of an empirical mass-luminosity relation was discovered (again unexpected), which could be applied to other AGN to estimate their black hole masses. As a result, it seems that the masses of the black holes in AGN follow the same relationship as in quiescent galaxies, which implies that during the growth of the MBH, the bulge is growing at about the same rhythm. 	

\bigskip 	
In conclusion, all this story tells us that Science evolves in an unpredictible way! Who could have thought in the sixties that massive black holes would invade our Universe and be present everywhere around us? It tells us also that, Science being made by human beings, it suffers from the same weaknesses, and like them, it progresses slowly but ineluctably.

\end{document}